\documentclass[fleqn,usenatbib]{mnras}
\usepackage{newtxtext,newtxmath}
\usepackage[T1]{fontenc}
\DeclareRobustCommand{\VAN}[3]{#2}
\let\VANthebibliography\thebibliography
\def\thebibliography{\DeclareRobustCommand{\VAN}[3]{##3}\VANthebibliography}

\usepackage{times,refname,bm}
\usepackage{natbib,url}
\usepackage{ctable,tabularx}
\usepackage{amstext,amsmath,multicol,verbatim,eurosym,graphicx,epsfig,latexsym}
\usepackage{caption}
\usepackage{booktabs} 
\bibpunct{(}{)}{;}{a}{}{,}

\newcommand{\be}{\begin{equation}}
\newcommand{\ee}{\end{equation}}
\newcommand{\bea}{\begin{eqnarray}}
\newcommand{\eea}{\end{eqnarray}}
\newcommand{\rund}[1]{\left(#1\right)}

\newcommand{\eck}[1]{\left[ #1 \right]}

\title[magnification bias]{\textbf{Bias in apparent dispersion measure due to de-magnification of plasma lensing on background radio sources}}

\author[Er et al.]{Xinzhong Er$^{1}$\thanks{Email: phioen@163.com}, Jiangchuan Yu$^2$, Adam Rogers$^3$, Shihang Liu$^2$, Shude Mao$^{4,5}$\\
$^1$South-Western Institute for Astronomy Research, Yunnan University, Kunming, Yunnan, 65000, P.R.China\\
$^2$Department of Astronomy, Yunnan University, Kunming, Yunnan, P.R.China\\
$^3$Department of Physics and Astronomy, University of Manitoba, Winnipeg, R3T 2N2, Canada\\
$^4$Department of Astronomy, Tsinghua University, 100084 Beijing, P.R.China\\
$^5$National Astronomical Observatories, Chinese Academy of Sciences, 20A Datun Road, Chaoyang District, Beijing 100101, P.R.China\\
}
\date{Accepted XXX. Received YYY; in original form ZZZ}
\pubyear{202x}

\begin{document}
\label{firstpage}
\pagerange{\pageref{firstpage}--\pageref{lastpage}}
\maketitle

\begin{abstract}
The effect of ionized gas on the propagation of radio signals is known as plasma lensing. Unlike gravitational lensing, plasma lensing causes both magnification and strong de-magnification effects to background sources. We study the cross section of plasma lensing for two density profiles, the Gaussian and power-law models. In general, the cross section increases with the density gradient. Radio sources can be used to measure the free electron density along the line of sight. However, plasma lensing de-magnification causes an underestimate of the electron density. Such a bias increases with the electron density, and can be up to $\sim 15\%$ in the high density region. There is a large probability that high density clumps will be missed due to this bias. The magnification of plasma lensing can also change the luminosity function of the background sources. The number density of sources on both the high and low luminosity ends can be overestimated due to this biasing effect.
\end{abstract}

\section{Introduction} 
\label{sect:intro}

Propagation of light in the universe can be affected by several phenomena, e.g. gravitational field generated by massive objects. An example of such phenomena is gravitational lensing, which provides a powerful tool in studying matter distributions throughout the Universe \citep[e.g.][]{2006glsw.book.....S}. In addition to gravitational lensing, inhomogeneous distributions of free electrons can cause deflection of light rays as well, called plasma lensing \citep[e.g.][]{CleggFL1998,Tuntsov2016,FRBplasma1}. Plasma lensing shares several similar features with gravitational lensing, especially in terms of the mathematical description. However, different features introduced by plasma lensing (such as wavelength dependence) become significant only at low frequency, e.g. usually in the radio band. Moreover, the deflection caused by plasma lensing is opposite to the deflection due to gravitational lensing, which makes an over-dense clump of plasma to act as a diverging lens. 
Observations of plasma lensing began with the discovery of wild changes in the flux density of compact radio sources \citep{ESE0}, a phenomenon known as Extreme Scattering Events (ESEs).  Although a detailed physical picture of the phenomenon remains elusive, it is believed that wavelength-dependent refraction due to plasma along the line of sight in an ESE cannot be neglected. Since the introduction of the Gaussian plasma lens model, it has been widely used to describe the phenomenon \citep{romani87,CleggFL1998}.

Millisecond duration pulses, known as fast radio bursts \citep{2007Sci...318..777L,frbreview2019,2019A&ARv..27....4P}, are dispersed during propagating in the ionised gas of the interstellar and intergalactic media. Several observational aspects of FRBs are modulated by plasma lensing, such as the dispersion relation and amplitude. The cosmological origin of FRBs provide a potential probe to study the total integrated column density of free electrons along the line of sight. Studies of the baryon fraction in the intergalactic medium using FRBs with redshift have been recently introduced in the literature \citep[e.g.][]{2020Natur.581..391M,2021arXiv210108005J}. However, the frequency-time delay relation (i.e. the Dispersion Measure, DM), is not an exact estimate of the projected electron density \citep[e.g.][]{2020arXiv200702886K,er+2020,2021ApJ...911..102O}. Moreover, it has been noted that plasma lensing is responsible for both magnification and strong de-magnification of background sources from both observations \citep[e.g.][]{CleggFL1998} and analytical models \citep[e.g.][]{Tuntsov2016,er&rogers18}. The diverging properties of plasma lensing deflects radio signals that would otherwise be received by us without a plasma clump along the line of sight. Therefore, the probability of detecting background sources is modified by the existence of plasma lenses. The distribution of the observed background sources will not be uniform, and will always appear biased toward the low density region of the plasma clumps. With such a biased tracer, one expects another systematic error in the estimation of the electron density in addition to all the other uncertainties already mentioned. In this study, we perform a toy simulation to evaluate the effect of such a systematic density bias, neglecting all the other uncertainties.
We briefly summarise the basics of plasma lensing in Sect.\,\ref{sect:basic} and discuss the lensing cross section for magnification and de-magnification in Sect.\,\ref{sec:csection}. A toy simulation to evaluate the bias in estimating the electron density is presented in Sect.\,\ref{sec:mag-effect}. Finally we discuss our results in Sect.\,\ref{sec:summary}. We adopt the standard $\Lambda$CDM cosmology with parameters based on the results from the Planck 2018 \citep{2020A&A...641A...6P}: $\Omega_\Lambda=0.6847$, $\Omega_m=0.315$, and Hubble constant $H_0=100h$\,km\,s$^{-1}$\,Mpc$^{-1}$ with $h=0.674$.

\section{The basics of plasma lensing}
\label{sect:basic}
The description of plasma lensing follows from gravitational lensing, e.g. \cite{2006glsw.book.....S}. One can also find details from, e.g., \citet[][]{Tuntsov2016,FRBplasma1,er&rogers18}. The lens is approximated as thin and the deflection angles are assumed to be small. We consider a source at angular position $\beta$ on the source plane with respect to the line of sight. The corresponding image of this source is then formed at the angular position $\theta$ on the observer's sky. For simplicity we only consider axisymmetric lens models. The lens equation can be thus related by the deflection angle $\alpha$
\begin{equation}
\beta=\theta-\alpha(\theta)\,=\theta - \nabla_\theta\psi (\theta),\qquad \theta=\sqrt{\theta_x^2+\theta_y^2},
\label{eq:lenseq}
\end{equation}
where $\psi$ is the effective lens potential, and $\nabla_\theta$ is the gradient on the image plane. In plasma lensing, the effective potential is determined by the projected electron density $N_{\rm e}(\theta)$, and is given by
\begin{equation}
\psi(\theta) \equiv \frac{D_{ds}}{D_d D_s} \frac{\lambda^2}{2\pi} r_e\, N_{\rm e}(\theta),
\end{equation}
where $D_{ds}$, $D_d$, and $D_s$ are the angular diameter distance from lens to source, from lens to observer, and from observer to source respectively. $r_e$ is the classical electron radius, and the wavelength of the signal is $\lambda$. In principle, the observed wavelength is redshifted due to the expansion of the universe, i.e. $\lambda^{\rm obs}=(1+z)\lambda$. Since we only consider lenses at low redshifts, such an effect is tiny and will be neglected.  In this work, we mainly focus on point-like sources, such as pulsars or FRBs. Due to the weakness of the lensing effects, image distortions cannot be observed directly. Moreover, the phases of radio signals are not included in the current study, which may affect observations of coherent sources. Thus, the magnification and frequency-dependent time-delay are the main observables. Plasma lensing usually produces behaviour analogous to a diverging lens, and causes significant de-magnification, i.e. $|\mu|\ll 1$. Magnification can be calculated from the Jacobian $A$ of the lens equation (\ref{eq:lenseq}), such that $\mu^{-1}={\rm det(A)}$. In the axi-symmetric lens, the magnification simplifies,
\be
\mu^{-1}={\beta \over \theta} {d \beta \over d \theta}.
\ee

Two analytical toy models are adopted to describe the number density of electrons. The first is the widely used Gaussian model \citep[e.g.][]{CleggFL1998,FRBplasma1}, which can also serve as a building block for constructing more complicated density profiles. The second profile we consider is a power-law model, of which the lensing properties can be calculated analytically as well. Moreover, the maximum density gradient of the  power-law lens model, which depends on the power-law index, can be much higher than that of a Gaussian profile even with a low average density. It is thus interesting to explore both profiles as they can be adopted for different scenarios. A Gaussian random field model provides a good description of the Kolmogorov spectrum of electron densities \citep[e.g.][]{2016ApJ...817...16C}, which will be studied in a future work.

\subsection{Magnification profiles}
\begin{figure*}
\centerline{
\includegraphics[width=5.8cm]{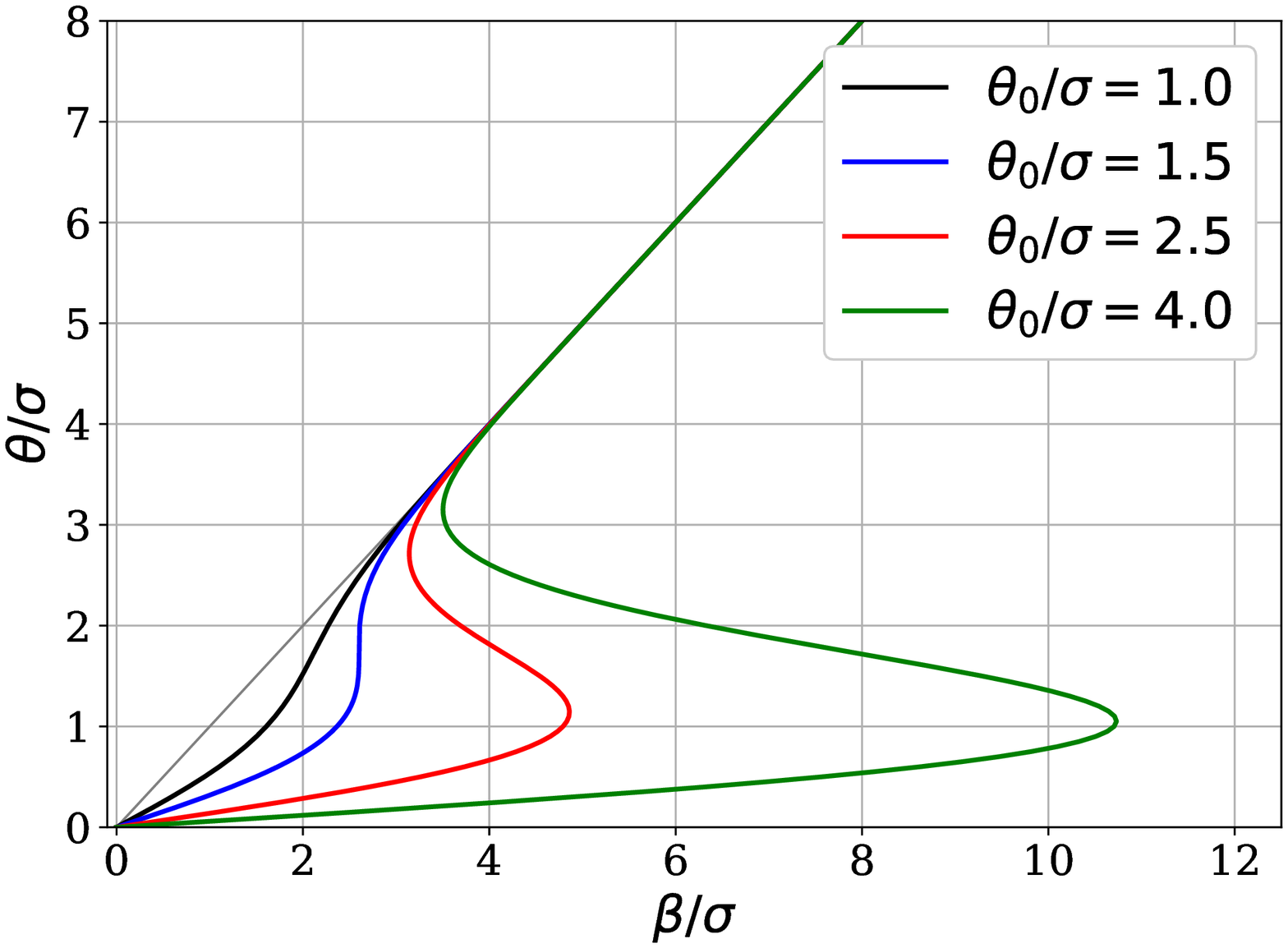}
\includegraphics[width=5.8cm]{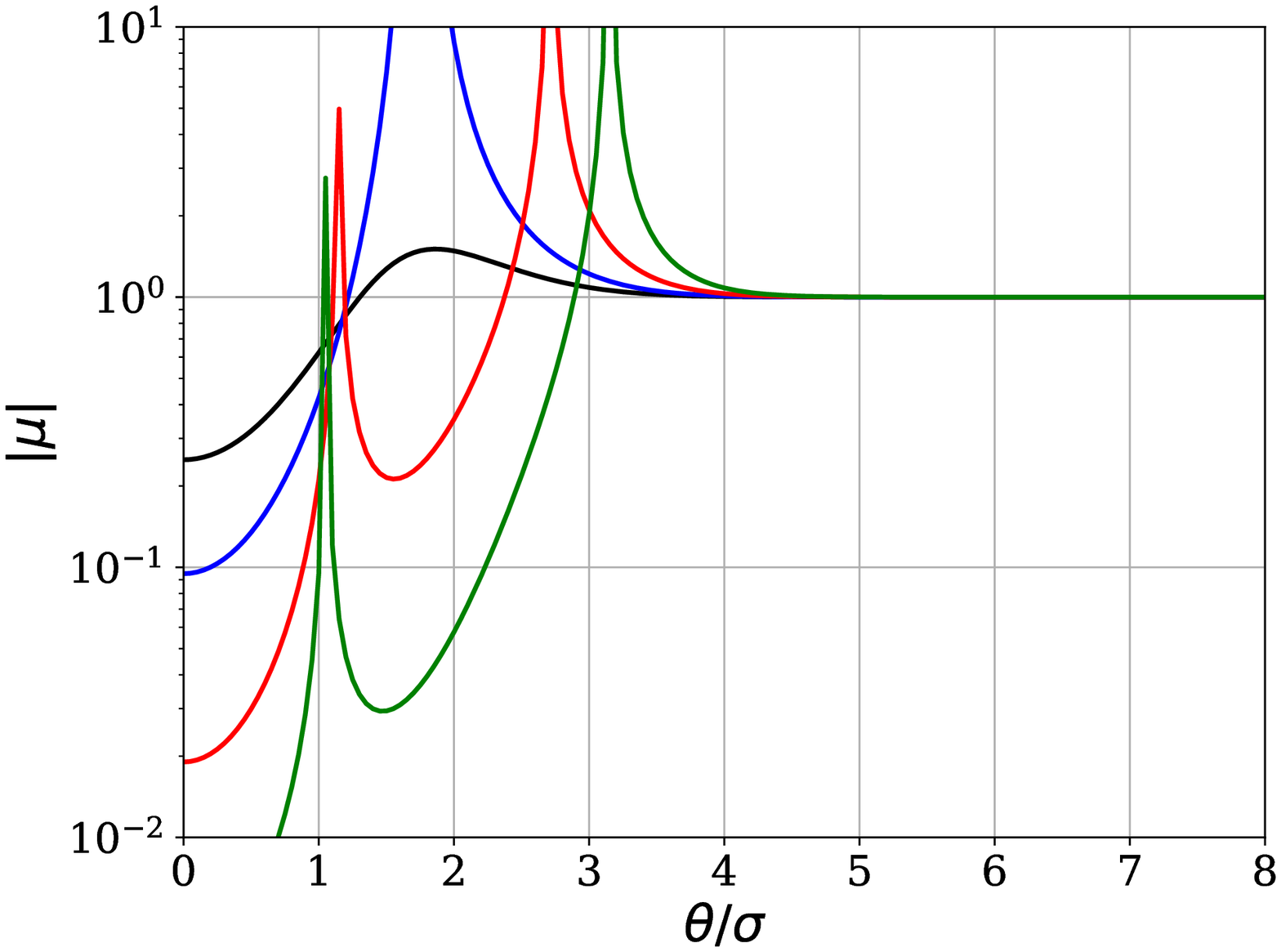}
\includegraphics[width=5.8cm]{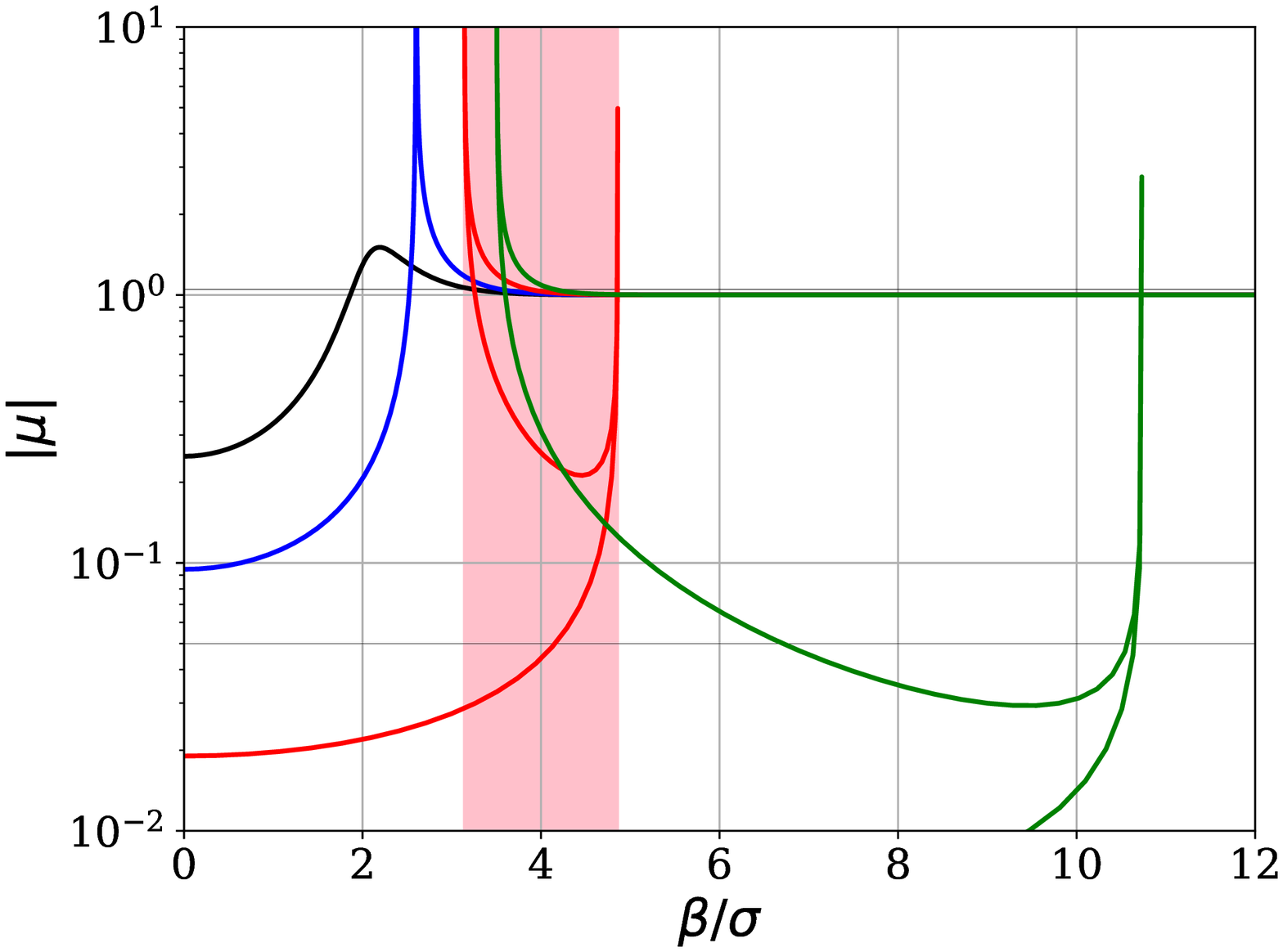}}
\caption{Lensing properties of the Gaussian model:the $\theta-\beta$ relation (left), magnification curve on the image plane (middle), and magnification curve on the source plane (right), respectively. The pink shaded region in the right panel shows the cross section where three images are generated for $\theta_0/\sigma=2.5$ (red curves). The equivalent parameter ratios for the SPL lens using Eq.\,\ref{eq:gaussian2spl} are $\theta_0/\theta_c=1.4,1.8,2.5,3.4$ respectively. }
\label{fig:gaussian1}
\end{figure*}
\begin{figure*}
\centerline{\includegraphics[width=5.8cm]{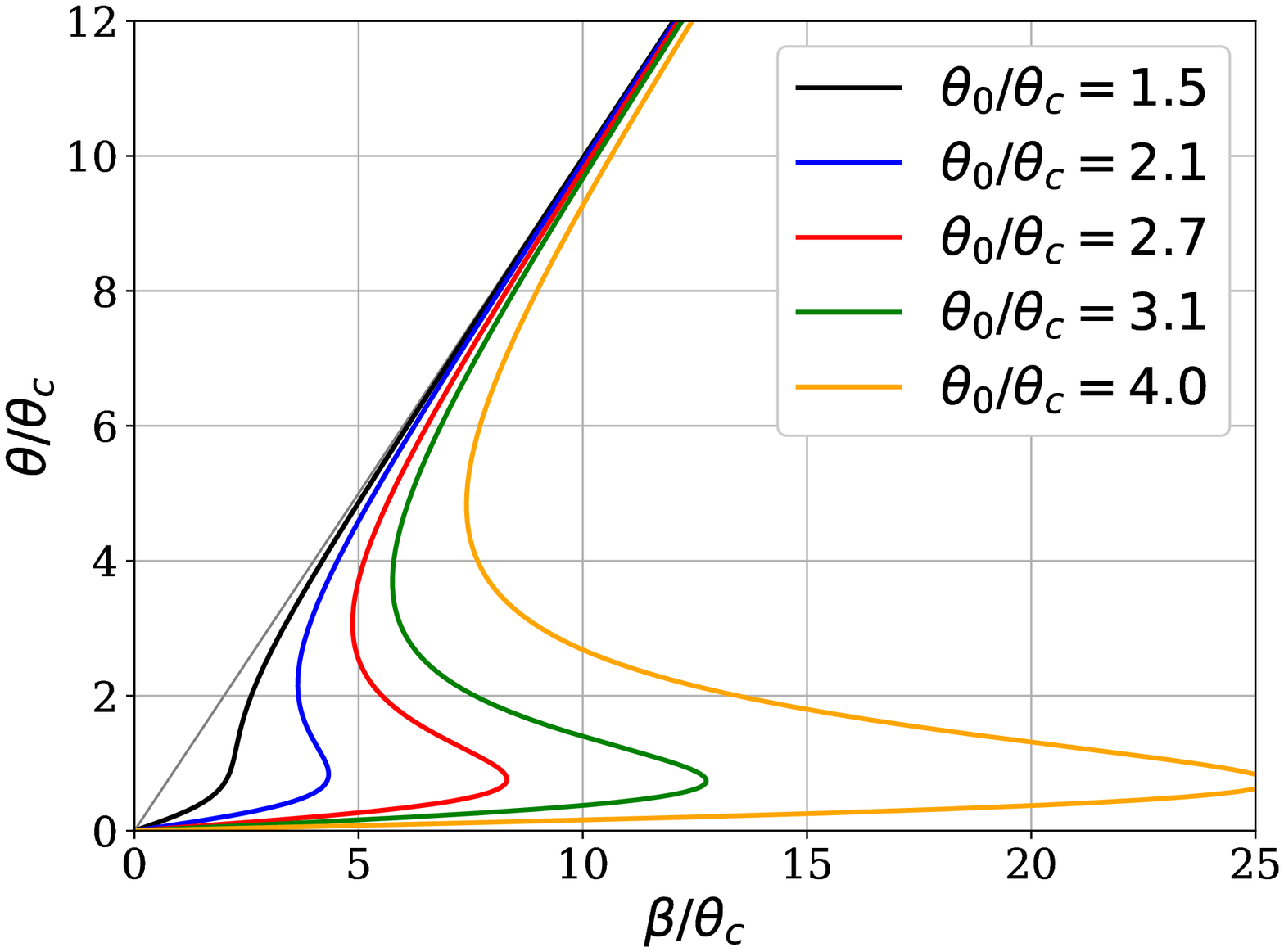}
\includegraphics[width=5.8cm]{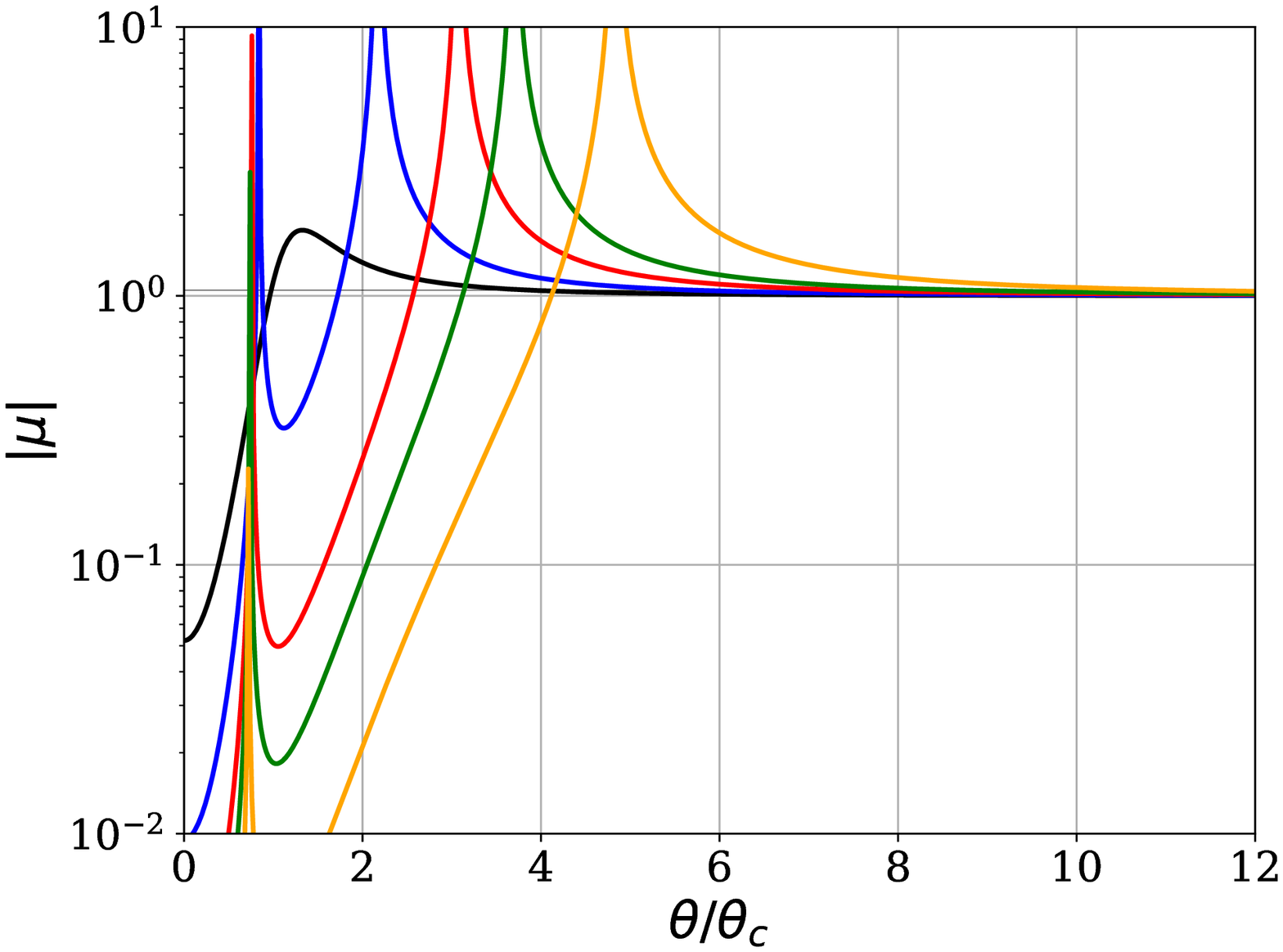}
\includegraphics[width=5.8cm]{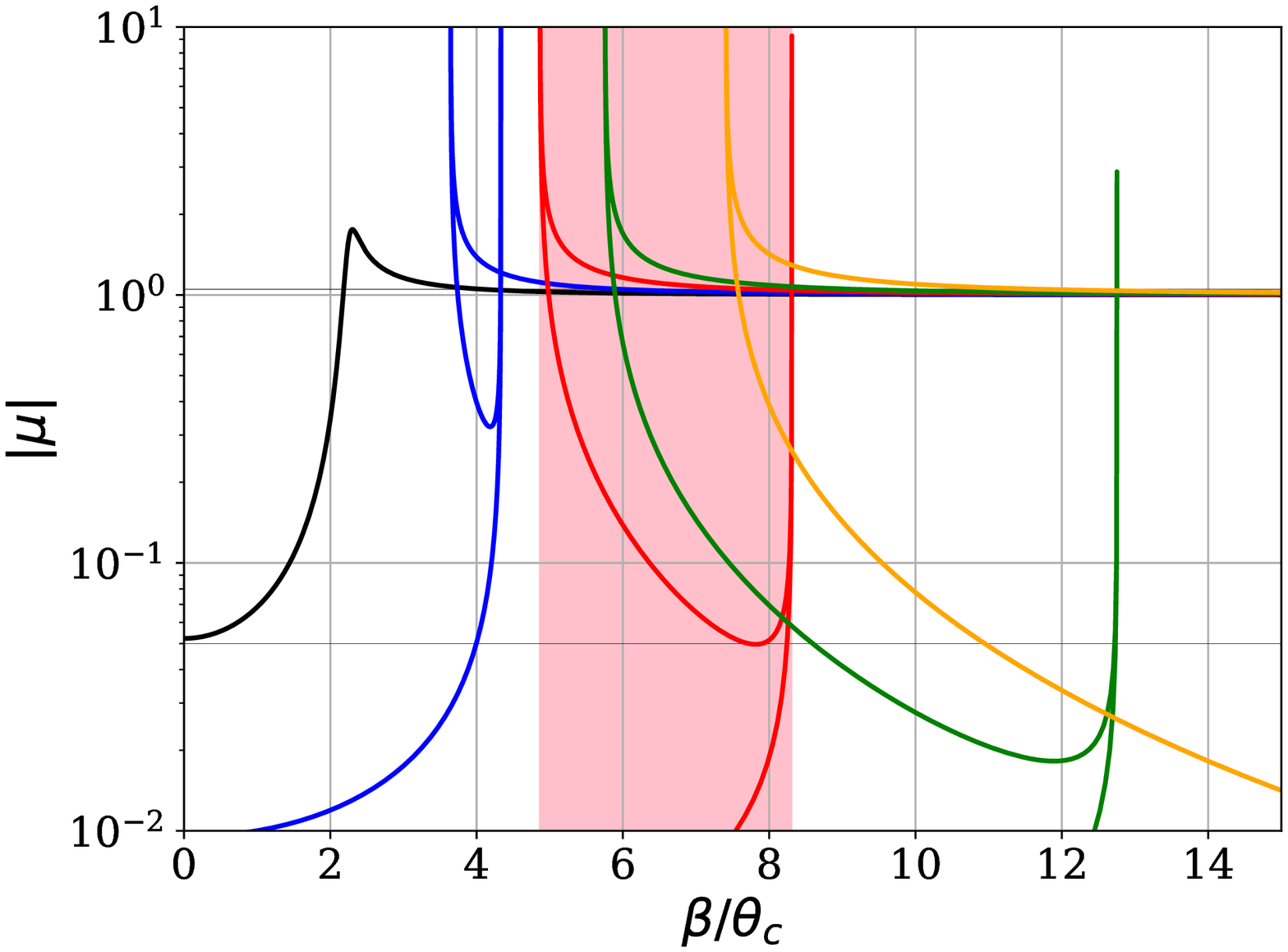}}
\caption{Lensing properties of the softened power-law model: the $\theta-\beta$ relation (left), and the magnifications curves on the image (middle) and source planes (right), respectively. The equivalent parameter ratios for the Gaussian lens using Eq.\,\ref{eq:gaussian2spl} are $\theta_0/\sigma=1.2, 1.9, 2.8, 3.5, 5.1$ respectively. }
\label{fig:powerlaw1}
\end{figure*}

\subsubsection{A Gaussian model}
Let us first consider a Gaussian model centred at the origin of the coordinate system as our example lens. The electron column density can be given by
\begin{equation}
N_{\rm e}(\theta)=N_0 {\rm exp}\rund{-\dfrac{\theta^2}{2\sigma^2}},\quad (\theta>0),
\end{equation}
with $N_0$ the maximum electron column density within the lens and $\sigma$ as the width of the lens. The density gradient reaches its maximum at $\theta=\sigma$, which is given by
\be
N_{\rm e}'(\theta=\sigma)=-\frac{N_0}{\sigma} {\rm exp}\rund{-0.5}.
\label{eq:gaussian-gradient}
\ee
With this density profile, the effective lensing potential is 
\begin{equation}
\psi(\theta) = \theta_0^2 {\rm exp}\rund{-\dfrac{\theta^2}{2\sigma^2}},
\label{eq:gaussian-potential}
\end{equation}
with the angular scale given by
\be
\theta_0^2 = \lambda^2 \dfrac{D_{ds}}{D_s D_d}\frac{1}{2\pi} r_e N_0.
\ee
The inverse magnification can be written as
\be
\mu^{-1}(\theta) = 1+ {2\theta_0^2 \over \sigma^2} \rund{1-\dfrac{\theta^2}{2\sigma^2}} {\rm e}^{-{\theta^2\over 2\sigma^2}} + {\theta_0^4 \over \sigma^4} \rund{1-\dfrac{\theta^2}{\sigma^2}} {\rm e}^{-{\theta^2 \over 2\sigma^2}}.
\label{eq:mu-gaussian}
\ee
We see from Eq.\,\ref{eq:mu-gaussian} that the lensing properties of the Gaussian, such as the magnification and number of critical curves, depend on the ratio $\theta_0/\sigma$. For example, multiple images and caustics appear when $\theta_0>\theta_{\rm crit}$, where $\theta_{\rm crit}=\sqrt{{\rm e}^{3/2}/2} \sigma\approx 1.5 \sigma$ \citep{er&rogers18}. We evaluate the ratio $\theta_0/\sigma$, and in Fig.\,\ref{fig:gaussian1} show the image-source positions, as well as the magnification curves on the image and source planes, respectively. The image and source coordinates are normalised by the lens width $\sigma$.
In the left panel, peaks and valleys of all the curves appear at similar image positions, and they all approach the identity $\beta=\theta$ after about $\theta=4\sigma$. That gives similar radial positions for the critical curves, where the magnification curves of super-critical lenses peak around $\sim\sigma$ and $3\sigma$ (middle panel). In the right panel, the magnification as a function of the source position is shown. In the region where the curve corresponds to more than one magnification, multiple imaging occurs, e.g. in the pink shade region, three images are generated by the red curve lens. 

For a super critical lens (i.e. more than one caustic is generated by the lens), inside the inner caustic, only one image is formed by the lens with extremely de-magnification ($\mu\ll 1$). Thus once a source is located inside the inner caustic, we have low probability to detect any signal, and we call this region as ``forbidden''. As $\theta_0/\sigma$ increases, the inner caustic (which indicates the forbidden region) slowly moves outwards. In contrast, the position of the outer caustic moves outwards more rapidly, and the area between the two caustics grows. However, this does not have strong significance, because at large radii, the magnifications of the first image is close to unity, and those of the second and third images are extremely small (i.e. the images are strongly de-magnified and have a low probability of being observed). One can see this from the magnification curves in the pink shaded region in Fig.\,\ref{fig:gaussian1}.
In fact, the possible positions to detect two images is around the inner caustic, where the second image is only slightly de-magnified.

\subsubsection{A power-law model}
We now consider the power-law density model. Its three-dimensional electron density distribution is given by
\be
n_e( r) = n_0 {R_0^h \over r^h},
\ee
where $h>0$, and $n_0$ is a constant that represents the electron density at a constant characteristic radius $r=R_0$. The deflection angle of the power-law lens can be calculated analytically \citep{2009GrCo...15...20B,2010MNRAS.404.1790B,BTreview15}
\be
\alpha (\theta) = - {\theta_0^{h+1} \over \theta^h},
\ee
where the characteristic angular scale is
\be
\theta_0 = \rund{\lambda^2 \dfrac{D_{ds}}{D_sD_d^h} \dfrac{r_en_0 R_0^h}{\sqrt{\pi}} \dfrac{\Gamma({h+1 \over 2})}{\Gamma(h/2)}}^{1\over h+1}.
\ee
The electron density of the power-law model becomes infinite at the centre of the lens. In order to avoid this shortcoming, a finite core can be included in the density profile, with angular core radius $\theta_c$. Then one can simply add the core size in quadrature in the usual expression for the radius $\theta\longrightarrow \sqrt{\theta^2+\theta_c^2}$. The deflection angle becomes
\be
\alpha(\theta) = - \theta_0^{h+1} \dfrac{\theta}{(\theta^2+\theta_c^2)^{h+1\over 2}}.
\ee
More details on the Softened Power-Law (SPL) plasma lensing model can be found in e.g. \citet[][]{er&rogers18,rogers&er19}. The power-law density profiles usually have a higher density gradient than the Gaussian profile, and the density gradient increases with the power index. We adopt the profile with $h=2$ in our study. The effective lens potential is
\be
\psi(\theta)=\dfrac{\theta_0^3}{\sqrt{\theta^2+\theta_c^2}}.
\label{eq:spl-potential}
\ee
The density gradient of the SPL lens with $h=2$ reaches its maximum at $\theta=\theta_c/\sqrt{2}$, which is given by
\be
N'_{\rm e} (\theta=\theta_c/\sqrt{2})=-\frac{4\pi N_{\rm s0}}{3^{3/2}\theta_c^2}
\quad {\rm with}\quad 
N_{\rm s0}= \frac{n_0R_0^2}{D_d\sqrt{\pi}}\frac{\Gamma(3/2)}{\Gamma(1)}.
\label{eq:spl-gradient}
\ee
The inverse magnification is given by
\be
\mu^{-1} = 1+ \dfrac{\theta_0^3(2\theta_c^2-\theta^2)}{\rund{\theta^2+\theta_c^2}^{5/2}} + \dfrac{(\theta_c^2-2\theta^2)\theta_0^6}{\rund{\theta^2+\theta_c^2}^4}.
\label{eq:mu-powerlaw}
\ee
Analogous to the Gaussian lens, the magnification of the SPL lens profiles are also determined by the ratio of two parameters $\theta_0/\theta_c$, e.g. a critical curve appears when $\theta_0>\theta_c/\eck{2(2/5)^{5/2}}^{1/3}$. We show the lensing of the SPL model in Fig.\,\ref{fig:powerlaw1}. The forbidden region increases with $\theta_0/\theta_c$ as well. However, the similarities between the models end here. In fact the SPL lens shows several significant differences from the Gaussian lens. For power-law lenses, the deflection curves approach the identity $\theta=\beta$ at different speeds after the first peak, which depends on the $\theta_0/\theta_c$ ratio. On the image plane, the position of the first critical curve is close to $\sim \theta_c$, while the position of the second critical curve increases constantly with $\theta_0/\theta_c$. The most important difference is that the forbidden region on the source plane can increase to a large area, and we do not find an upper limit with larger $\theta_0/\theta_c$ in additional tests.

\subsubsection{A simple comparison between two profiles}
The two density profiles have different properties and parameters. In order to compare the two profiles fairly, two conditions are adopted on the projected density and density gradient, namely, the central density and the maximum gradient of these two profiles are kept the same (using Eqs.\,\ref{eq:gaussian-potential} and \ref{eq:spl-potential}, and  Eqs.\,\ref{eq:gaussian-gradient} and \ref{eq:spl-gradient}). With these two constraints, one can obtain
\begin{align}
3^{3/2} \theta_c & = 2\sigma {\rm e}^{0.5},\\
\theta_{0,p}^3 & = \theta_{0,g}^2\theta_c,
\end{align}
where $\theta_{0,p}$ and $\theta_{0,g}$ are the angular scale of the SPL and the Gaussian lens respectively. Therefore, a conversion factor for the parameter ratio of density property between the two profiles is given by 
\be
\dfrac{\theta_{0,g}}{\sigma}
=\dfrac{2{\rm e}^{0.5}}{3^{3/2}} \rund{\theta_{0,p}\over \theta_c}^{3/2}.
\label{eq:gaussian2spl}
\ee
With such constraints, the density profiles can be compared less arbitrarily. However, as we will see later that the lensing strength generated by the two profiles are dramatically different. In the caption of Figs.\,\ref{fig:gaussian1} and \ref{fig:powerlaw1}, the converted parameter ratios are given. One can see that the corresponding SPL lens will have similar lensing strength with the Gaussian lenses.

\begin{figure}
\centerline{\includegraphics[width=10cm]{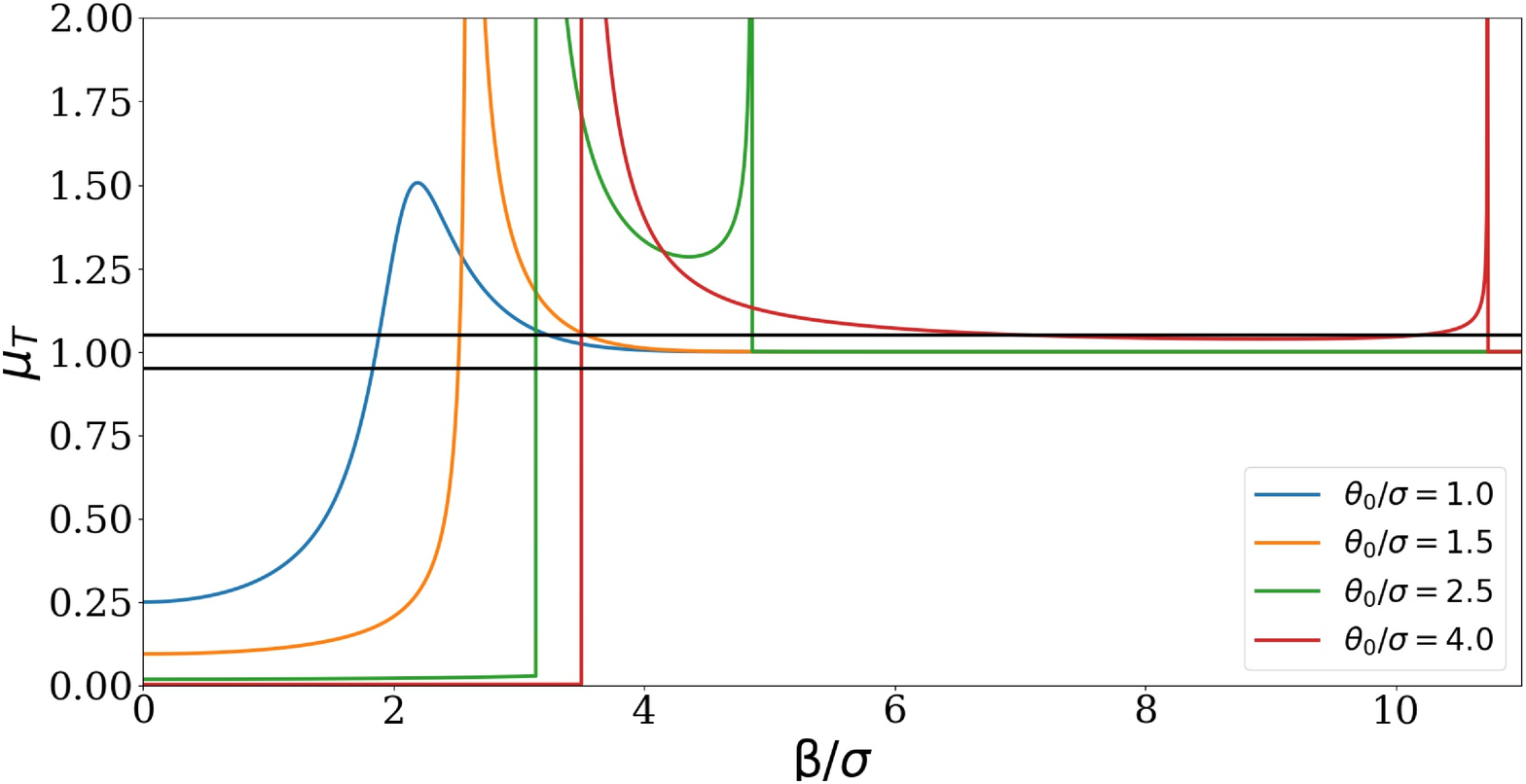}}
\centerline{\includegraphics[width=10cm]{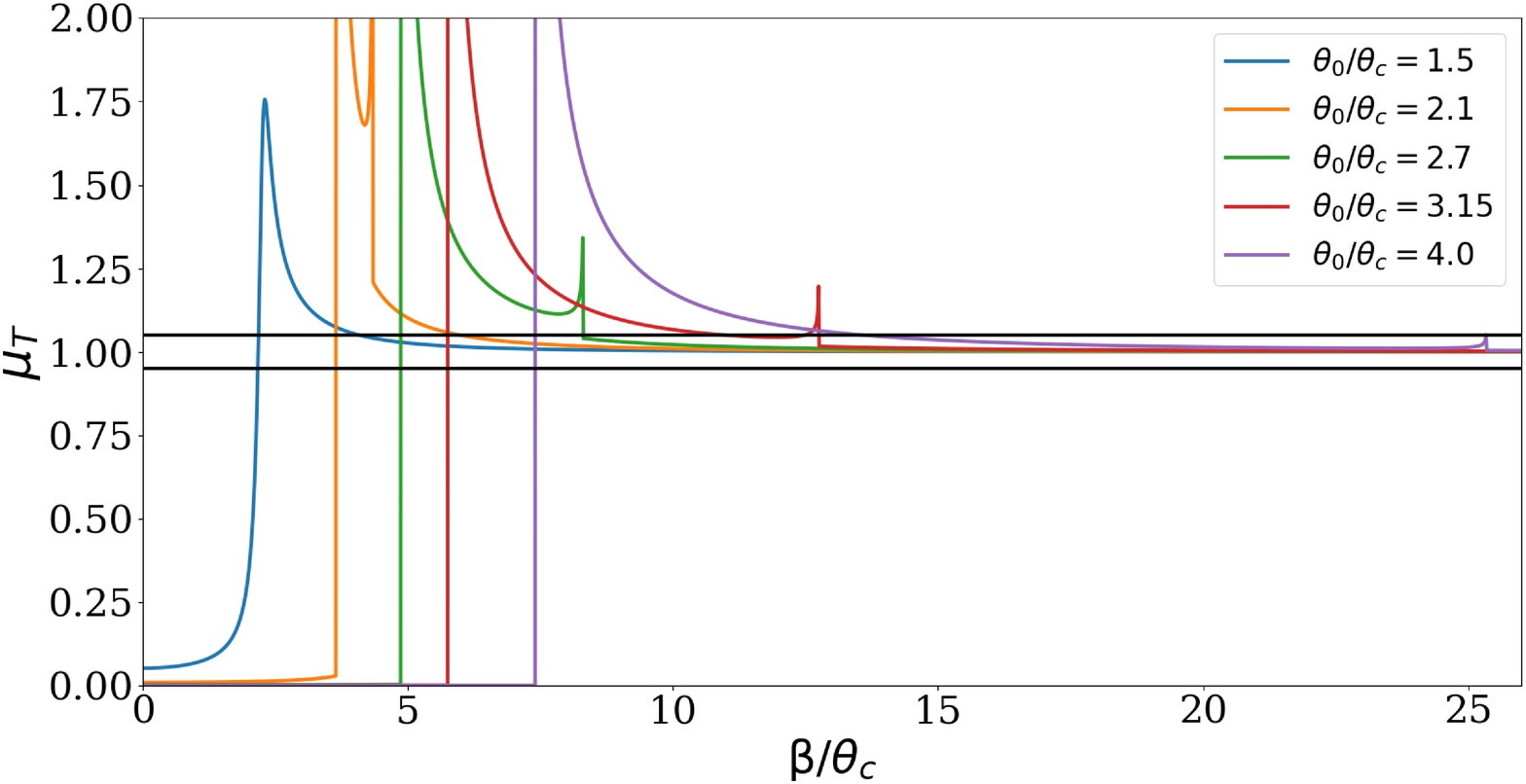}}
\caption{The magnification curves of the Gaussian lens for different $\theta_0/\sigma$s (top) and the softened power-law model for different $\theta_0/\theta_c$s. The two horizontal lines indicate the $\mu_{\rm T}=0.95,1.05,$ respectively.}
\label{fig:Gaussian_muT}
\end{figure}
\begin{figure}
\centerline{\includegraphics[width=8cm]{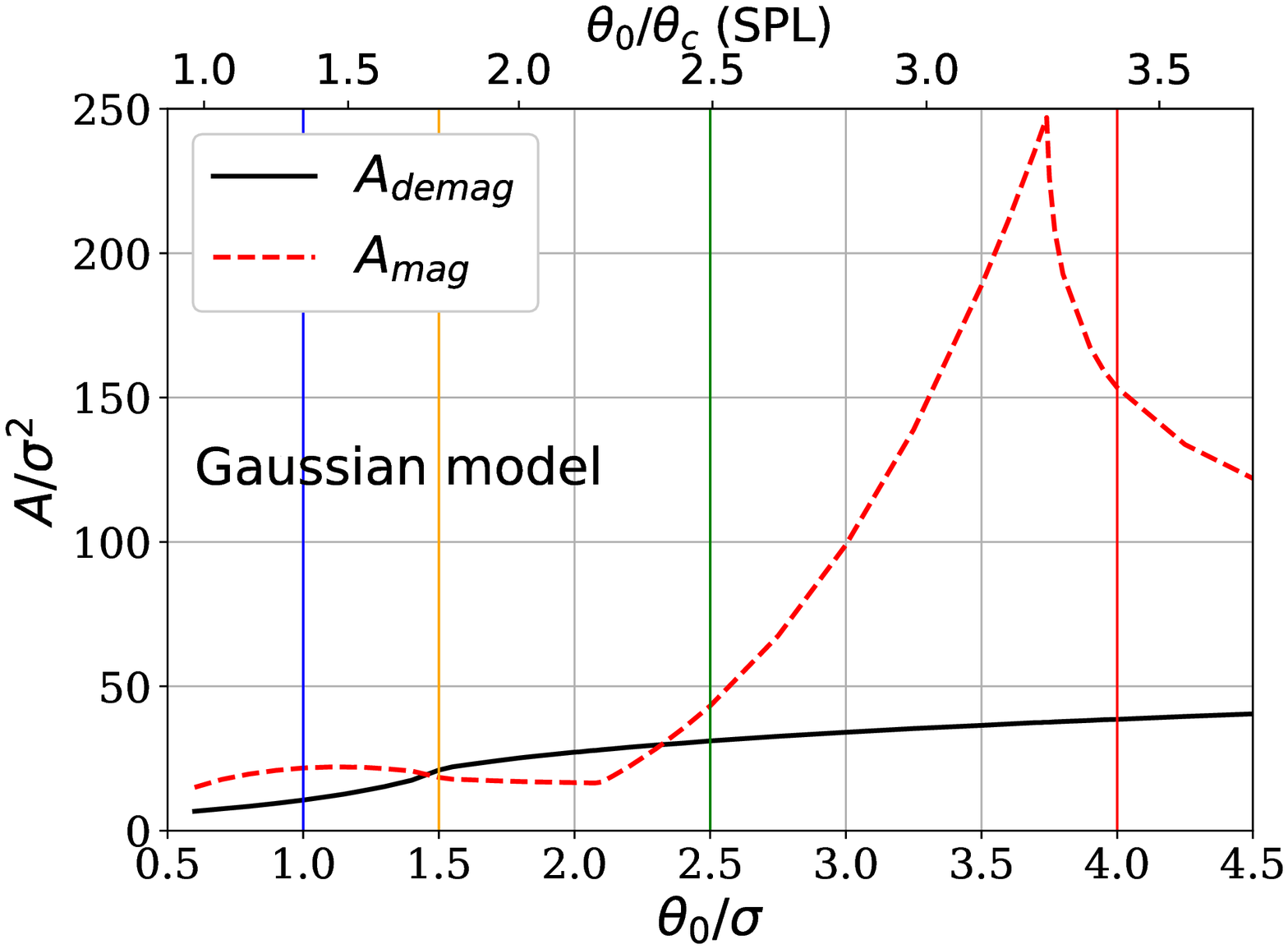}}
\centerline{\includegraphics[width=8cm]{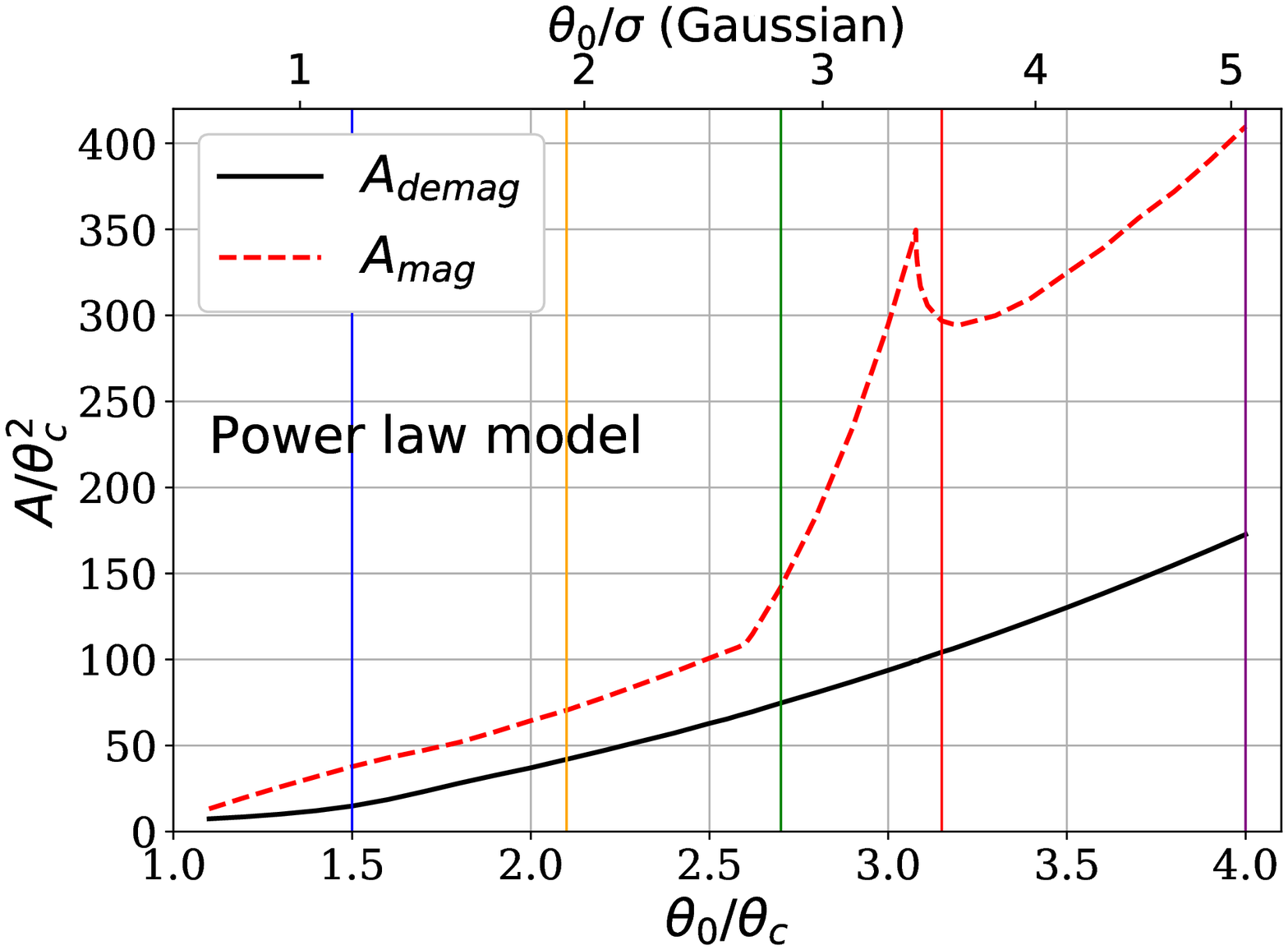}}
\caption{The magnification or de-magnification probabilities of plasma lens for different scenarios: $top$-Prob.\,vs.\,$\theta_0/\sigma$ for the Gaussian model, $bottom$-Prob.\,vs.\,$\theta_0/\theta_c$ for the power-law model. The colour vertical lines indicate the ratios of which the magnification curves are shown in Fig.\,\ref{fig:Gaussian_muT}. The corresponding parameter ratios of the other profile are shown at the top horizontal axis.}
\label{fig:2lens-prob}
\end{figure}
\section{Lensing Cross section}
\label{sec:csection}
Plasma lensing generates both magnification and de-magnification that affect background sources. It is interesting to compare the lensing cross section of the two effects. In plasma lensing, the weak deflection causes tiny separations between multiple images. There is only some evidence of resolved multiple images from observations with high resolution, \citep[e.g.][]{2000ApJ...543..740B,2010ApJ...708..232B}. We thus calculate the total magnification $\mu_T$ on the source plane, which is given by the sum of all the image magnifications produced by the lens
\be
\mu_{T} =\sum_{i} |\mu_i|,
\ee
where $i$ is the index of a particular image. Such a total magnification does not contain the phase information of the images, and thus may give slightly different estimates for coherent sources. We define the area of magnification ($\mu_{T}>1.05$) or de-magnification ($\mu_{T}<0.95$) region on the source plane as, $A_{\rm mag}$ and $A_{\rm demag}$. 
%

First, we present the magnification curves for the Gaussian and SPL models with different ratios in Fig.\,\ref{fig:Gaussian_muT}. As we see, the parameter ratio ($\theta_0/\sigma$ for Gaussian lens, and $\theta_0/\theta_c$ for SPL lens) fully determines the magnification properties of the lens. And it is the same to the lensing cross section.
For both profiles, the de-magnifications of the sub-critical lenses are mild. Once they become critical, they can generate a clear forbidden region. The cross section of magnification increases with the ratio and is mainly between the two caustics.
In Fig.\,\ref{fig:2lens-prob}, we show the lensing cross section as a function of the parameter ratio. For the Gaussian model, the rate of increase of the de-magnification becomes slightly slower after $\theta_{\rm crit}$, while for a SPL model, such a change does not appear. The magnification curve is somewhat complicated. A small transition appears at $\theta_{\rm crit}$ as well. The dramatic change appears after $\theta_0/\sigma\sim 2.1$, where the outer caustic moves outward rapidly. When the ratio becomes sufficiently large, i.e. $\theta_0/\sigma\sim 3.7$, the probability of magnification decreases. The reason is that in some area between the two peaks, the magnification is below 1.05 and will be excluded, e.g. see the red curve in Fig.\,\ref{fig:Gaussian_muT}. However, for a lens with high $\theta_0/\sigma$, the high probability of magnification does not suggest lensing strength as one can see that most of the magnification in the cross section is only slightly larger than unity.

For the SPL lens, the probability of both effects is similar as that of Gaussian lens, i.e. increases with the ratio as well. The slope of magnification curve is larger between $\theta_0/\theta_c\sim[2.6,3.1]$. After that, there is a small decrease. It is the same as that for the Gaussian model, in some area between the two peaks, the magnification is below 1.05. 
In order to compare the lensing efficiency of the two profiles, we show the corresponding parameter ratios of the other profile using Eq.\,\ref{eq:gaussian2spl} at the top x-axis in each panel of Fig.\,\ref{fig:2lens-prob}.

\begin{figure*}
\centerline{
\includegraphics[width=5.5cm]{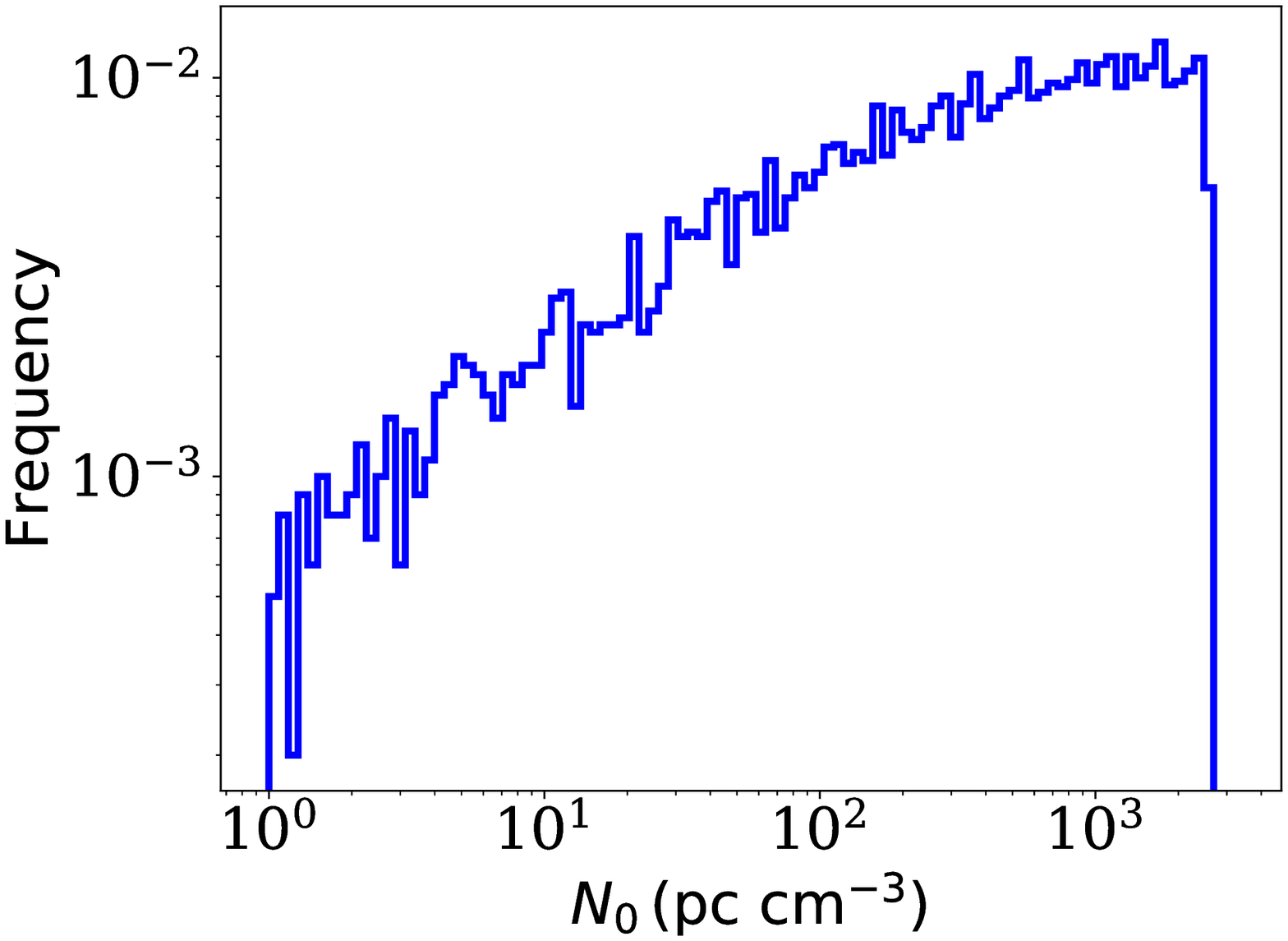}
\includegraphics[width=5.5cm]{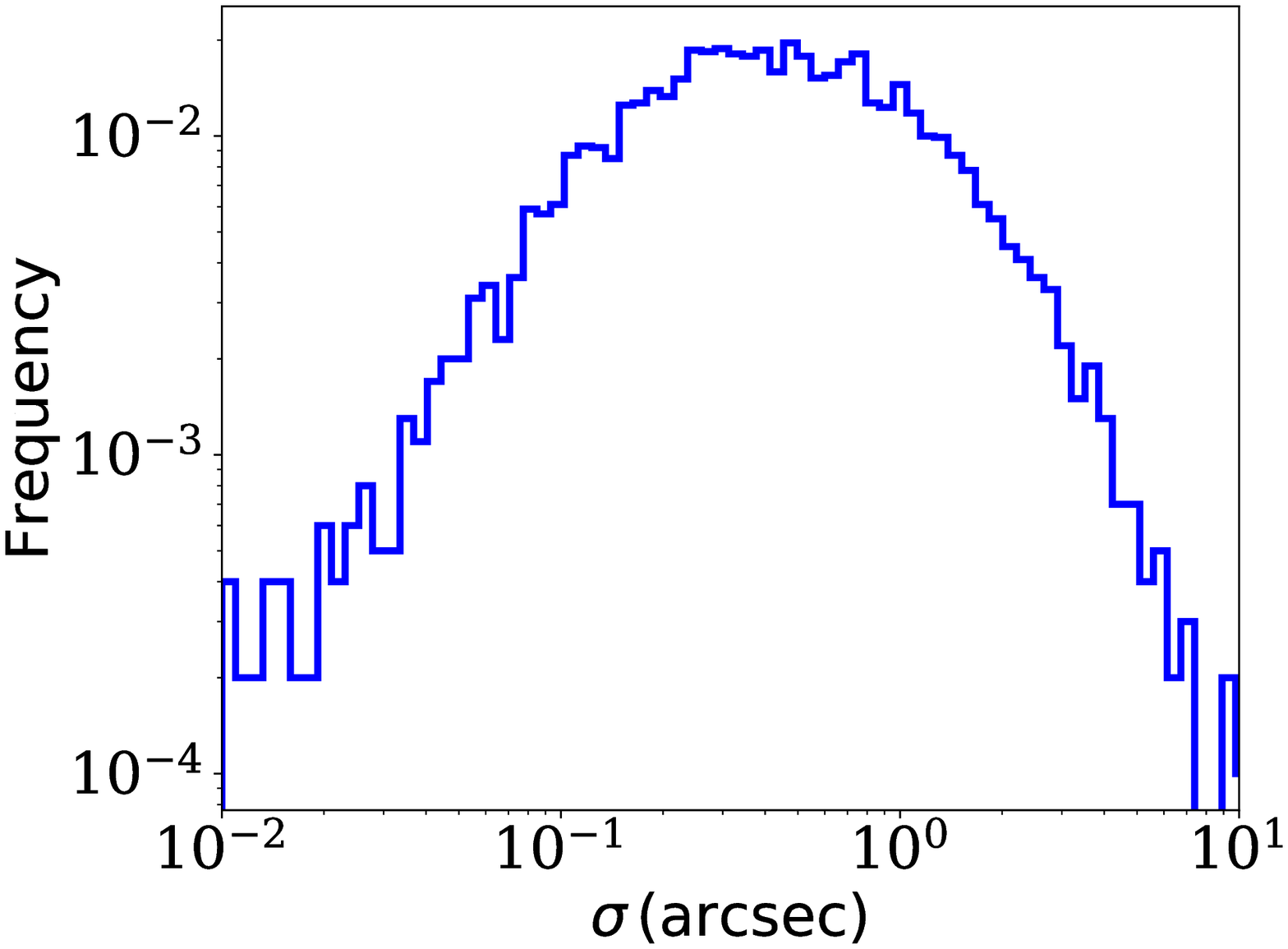}
\includegraphics[width=5.5cm]{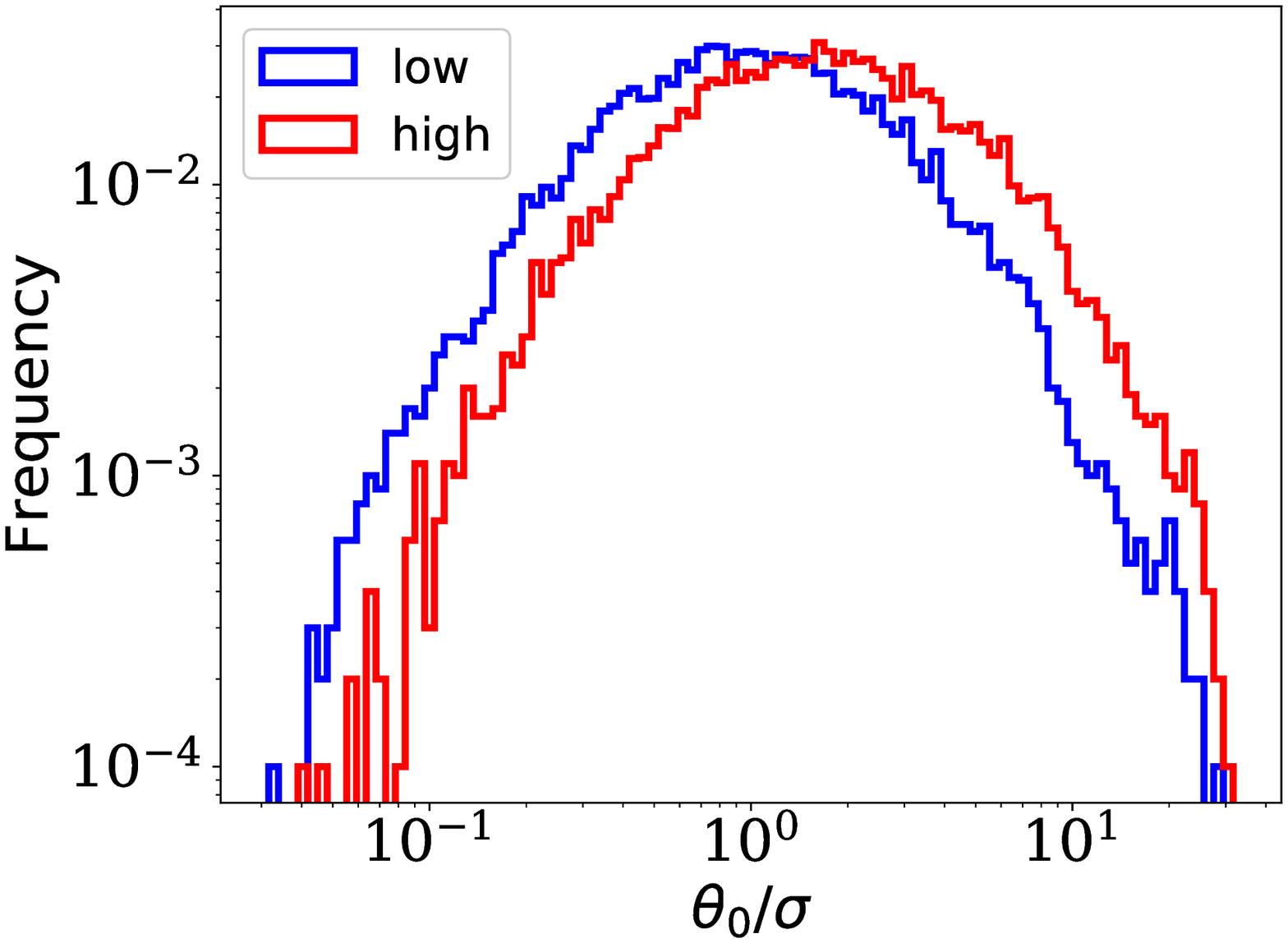}}
\caption{The distribution of lens parameters ($N_0,\, \sigma,\, \theta_0/\sigma$) that we used in our simulations. The red curve shows the distribution of the high density gradient case.}
\label{fig:lens-dist}
\end{figure*}
\begin{figure}
\includegraphics[width=8.5cm]{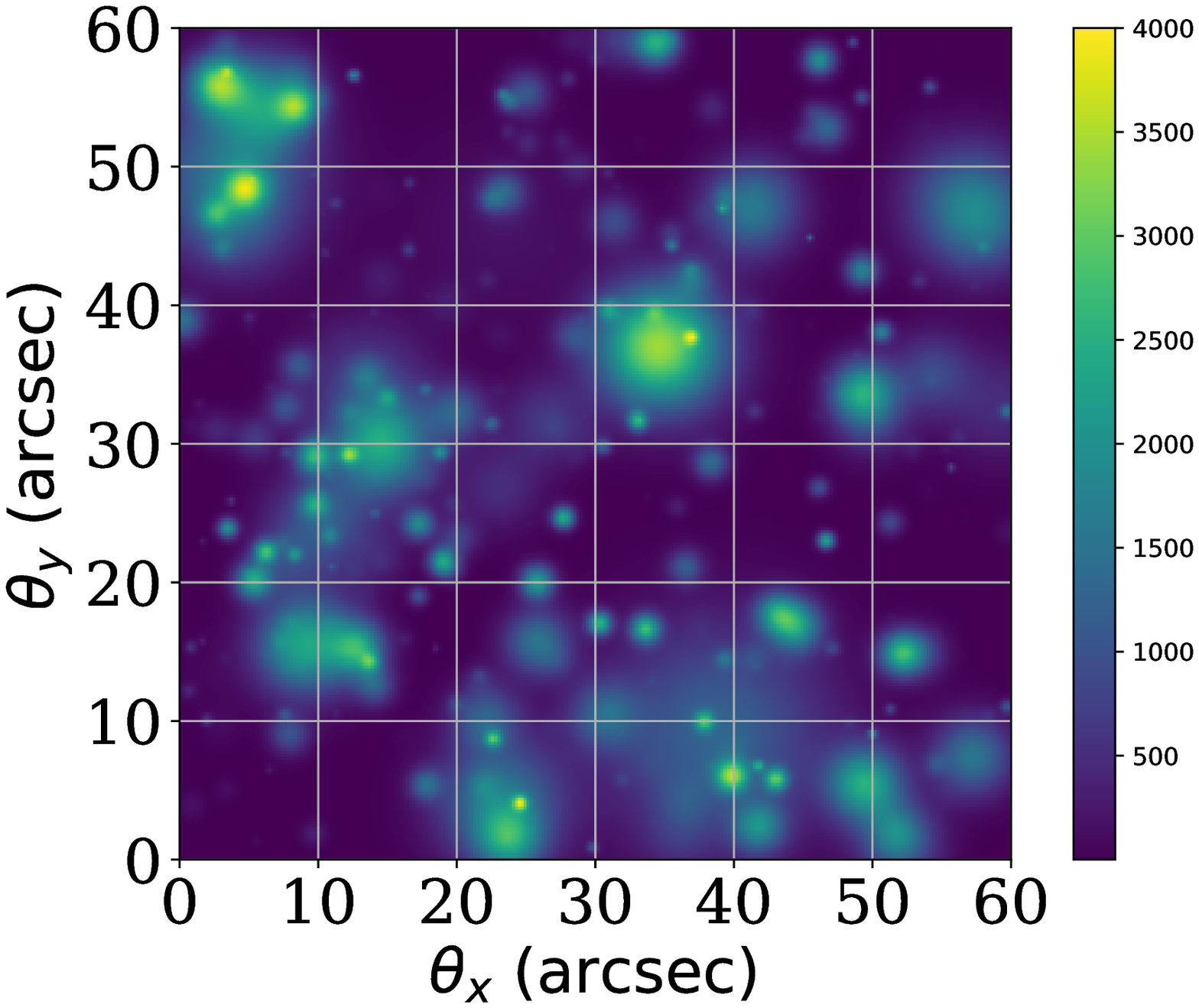}\\
\includegraphics[width=8.5cm]{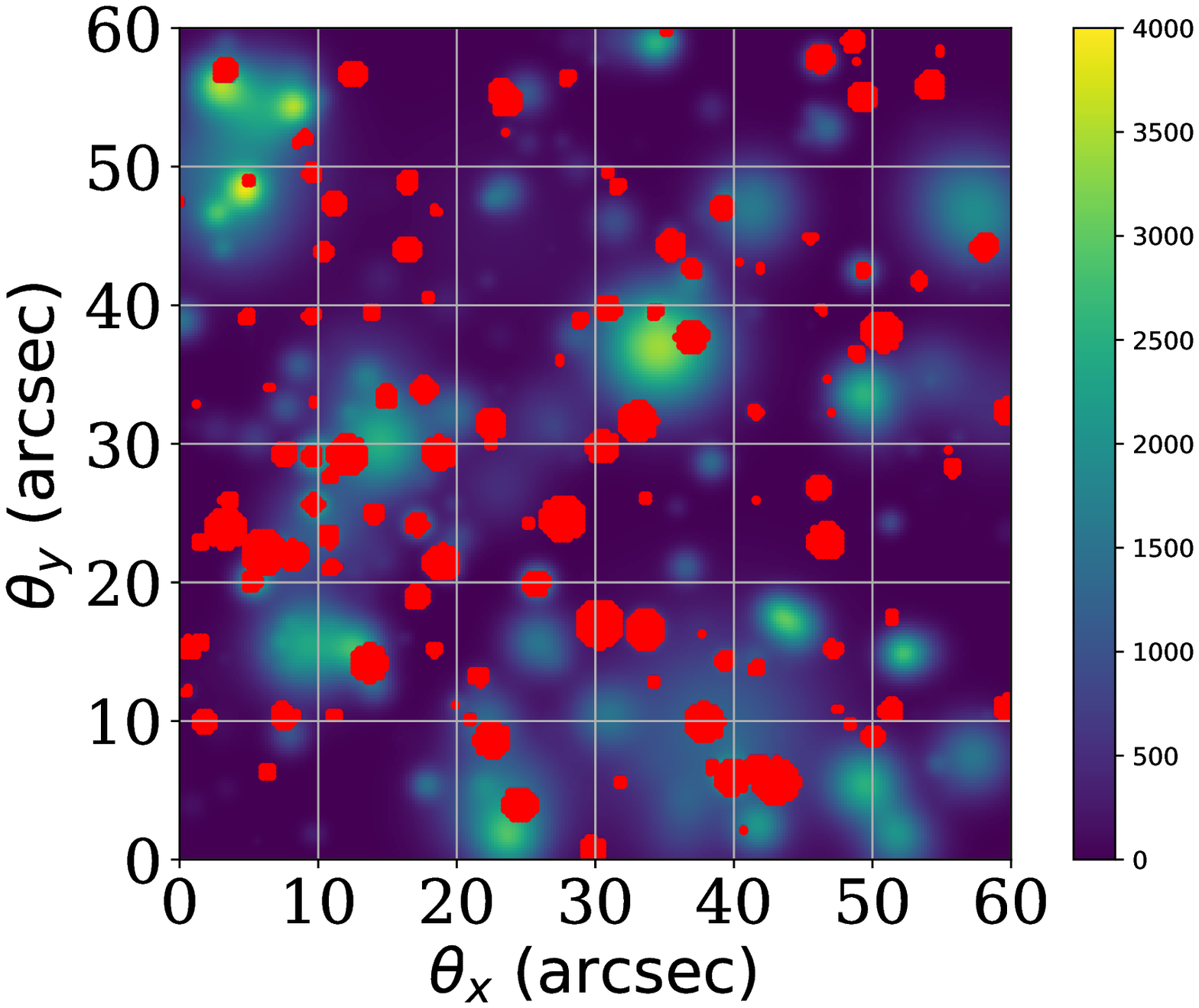}
\caption{One realisation of maps of the electron density over the field in our test: the top is the initial true map, while the bottom is the map excluding the de-magnification regions (red). The unit of the colour code is pc\,cm$^{-3}$. $700$ plasma clumps are uniformly distributed on the field. There is $5\%$ probability of a background source falling in the exclusion region.}
\label{fig:dm-map}
\end{figure}
\section{Magnification effects}
\label{sec:mag-effect}
It has been proposed to use radio pulsing sources to trace the ionised baryons throughout the universe. The relation of the Dispersion Measure to electron density, however, relies on several assumptions and approximations \citep[e.g.][]{2020arXiv200702886K}. For example, the contribution from ions is not included; the temperature of the plasma is assumed to be low, and the effect of magnetic field in plasma is not taken into account. Moreover, it is necessary to assume a uniform spatial distribution of the background radio sources. In gravitational lensing, the cosmic magnification effect can be used to study the matter distribution on large scales, but the inhomogeneous distribution of background sources has to be taken into account, \citep[e.g.][]{2005PhRvL..95x1302Z}. A significant difference of plasma lensing is that it can cause strong de-magnification to the source (even for the brightest primary image), which therefore cannot be observed at all. Thus, the information we extract from observations of radio sources will be in principle incomplete, a distinct conclusion compared to gravitational lensing. Such a ``selection effect'' can lead to a systematic bias in the estimate of the electron density. 

We perform a toy simulation to evaluate the magnification effect of plasma lensing. For simplicity, we adopt the redshift of a repeating FRB \citep[$z_s\approx0.2$,][]{repeatingFRB} for all of our sources. Additionally, we place all the plasma clumps in the Milky Way in order to avoid the redshift effect to the estimate of electron density \citep{2003ApJ...598L..79I,2004MNRAS.348..999I}. The observation frequency we adopted is $\nu=1$ GHz. A realistic model of the electron density on small scales is difficult to obtain, either from observations or simulations. We thus adopt a simple model and restrict the electron density in the range consistent with observations \citep[e.g.][]{NE2001a,NE2001b,2017ApJ...835...29Y}.

An axisymmetric Gaussian density model is used for all the plasma clumps in the tests in which $n_l$ lenses are randomly and uniformly placed in a field of $1\times1$ arcmin$^2$. The lens width $\sigma$ follows a log-normal distribution. The central density of the lens is sampled by two constraints: i) for each $\sigma$ the ratio $\theta_0/\sigma$ follows a log-normal distribution, the mean value of the distribution is $0.05$ if $\sigma$ is greater than 1 arcsec, and is $0.25$ if $\sigma$ is smaller than 1 arcsec, ii) $\theta_0<1$ arcsec to avoid an extremely large electron density (the corresponding $N_{0}$ is about $2600$ pc\,cm$^{-3}$). 

In Fig.\,\ref{fig:lens-dist}, we present the distributions of $N_0$, $\sigma$ and $\theta_0/\sigma$ of our mock lens samples. Most of the lenses have small density and size.
We calculate the magnification analytically or using finite difference. The plasma lens potential is calculated from the stacked electron density of all the lenses on a mesh with resolution of $0.24$ arcsec. The lensing properties, such as magnification can be performed by using the method in \citet{2005A&A...437...39B}. We regard the region where $\mu<0.25$ as ``unobservable'' on the lens plane, i.e. the deflection caused by the lens is not taken into account. The threshold of being unobservable depends on the condition of observations. In reality, $\mu\sim0.25$ is only a rough estimate of the forbidden regions. One can see from the figures of magnification curves (Fig.\,\ref{fig:Gaussian_muT}) that once the lens becomes critical, the magnification inside the caustic is extremely small around the lens centre. We also perform additional tests to calculate regions for $\mu<0.1$. The area of the unobservable regions decrease by only $10-20\%$, which is mainly caused by the sub-critical lenses. This will not change our conclusions substantially, but warrants a more detailed study with observational constraints in the future.

We mask out the unobservable regions on the lens plane, and obtain results to compare with those using all the simulated data.
In Fig.\,\ref{fig:dm-map}, we present one example of the density map of $N_{\rm e}$. One is the initial simulated map of the electron density, the other is the map we mark out the exclusion regions, which is shown by red colour. Most of the exclusion regions are located at high density region. But not all the high density regions are strongly de-magnified, since the gradient of density is another important factor as well. 

\begin{figure}
\includegraphics[width=8cm]{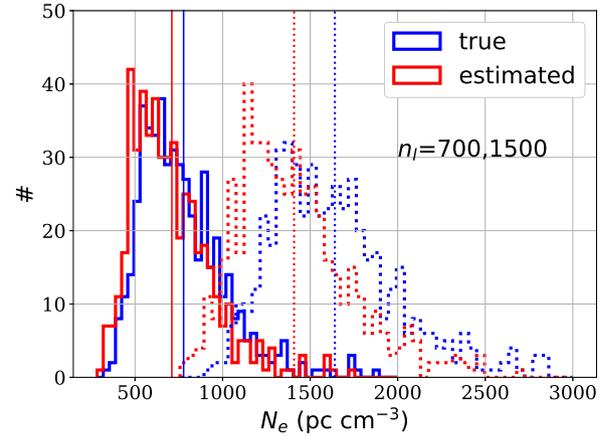}
\caption{The histogram of the average Dispersion Measure (DM) over the field for 500 realisations: the blue curve presents that over the whole field, the red curve excludes the de-magnification region ($\mu_{T}<0.25$). The solid (dotted) lines present that 700 (1500) plasma clumps are placed in the field. The vertical lines indicate the mean electron density $\bar{N_{\rm e}}$ over the 500 realisations.}
\label{fig:dm-hist}
\end{figure}

\begin{figure}
\includegraphics[width=8cm]{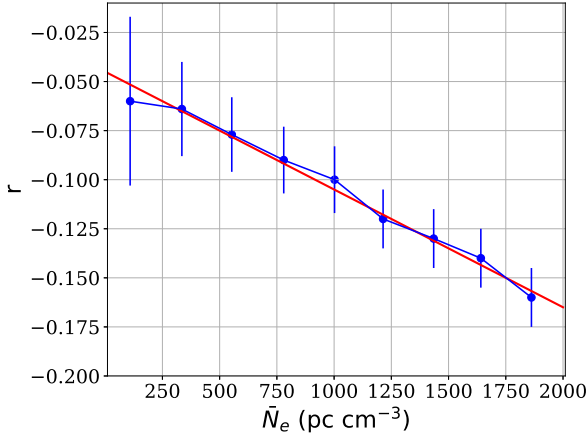}
\caption{The relative ratio (Eq.\,\ref{eq:dmratio}) as a function of $\bar{N_{\rm e}}$: the dots are calculated from our simulation. The red curve presents Eq.\,\ref{eq:r-dm-ana}.}
\label{fig:dm-underest}
\end{figure}

We compare the average electron density over the whole field (label as true) and over the field without the exclusion regions (label as estimated). 500 realisations are simulated to show the difference in Fig.\,\ref{fig:dm-hist}. The distributions with or without the exclusion regions show difference, which becomes significant with the increase of the number of lenses. With a higher electron density, the cross section of the exclusion region grows, and the probability of detecting a lower electron density becomes greater. We perform comparisons for different number of lenses $n_l$. A probability of a background source falling in the exclusion region is defined as
\be
\tau=A_{\rm ex}/A_{\rm tot},
\label{eq:tau-area}
\ee
where $A_{\rm ex}$ is the area of the exclusion region, and $A_{\rm tot}$ is the total field area. The probability increases from $\tau=5\%$ for $n_l=700$, to $11\%$ for $n_l=1500$ in our tests. 
Moreover, we calculate the mean of average electron density over $500$ realisations, and compare the true value with the estimated one by the relative ratio
\be
r\equiv (\bar{N}_{\rm e,est} - \bar{N}_{\rm e,true})/\bar{N}_{\rm e,true}.
\label{eq:dmratio}
\ee
We show the relative difference in Table\,\ref{tab:table1}. From both Fig.\,\ref{fig:dm-hist} and Table\,\ref{tab:table1}, one can see that the ``estimated'' density from mock observations will be lower than the ``real'' one. With an increase in the electron density (i.e., the number of lenses in front of the field), the underestimate becomes larger.
As one expects that the density gradient is another important aspect, we perform additional simulations with higher gradient (the mean value of  $\theta_0/\sigma$ changes to $0.05$, $0.5$ of the two log-normal distributions in the simulation, and the distribution of $\theta_0/\sigma$ is shown by the red curve in Fig.\,\ref{fig:lens-dist}). The comparison is presented by the bottom three rows in Table.\,\ref{tab:table1}. The underestimate becomes slightly stronger with the similar electron density. In Fig.\,\ref{fig:dm-underest}, we present $r$ of $\bar{N_{\rm e}}$, $r$ is estimated by the average and the error bars present the standard deviation over 500 realisations. As a comparison, we show a linear relation between $r$ and $\bar{N_{\rm e}}$ by the red curve
\be
r=-a\,\bar{N_{\rm e}}-b,
\label{eq:r-dm-ana}
\ee
where $a=6\times10^{-5}$ cm$^3$/pc, and $b=0.045$.
Our test (dots) follows the trend of Eq.\,\ref{eq:r-dm-ana}. Given the high dispersion measure of distant FRBs (a few thousands), we expect the underestimate of electron density due to de-magnification can be $9\%$ for $n_l=700$ (for $\tau\approx 5\%$), and up to $\sim15\%$ for $n_l=1500$ (for $\tau\approx 10\%$). Such a relation (Eq.\,\ref{eq:r-dm-ana}) only provides a rough estimate of the bias, since it depends on several factors of the density model, e.g. the distribution of $\theta_0/\sigma$ of the Gaussian lenses. It is clear that the relation will not be strictly valid for other density models.
More constraints from observations and further studies are necessary in order to obtain a more realistic estimate of the bias.

%
%

\begin{table}
  \begin{tabular}{c|c|c|c|c}
    \specialrule{.1em}{.05em}{.05em}
    $n_l$  &$\bar{N}_{\rm e,true}$  &$\bar{N}_{\rm e,est}$  &$r$ &$\tau$ \\
    \hline
    \hline
    \rule{0pt}{3ex}
    100  &109.6  &105.1  &-0.060 &0.0079\\
    700 &780.1 &712.8 &-0.086 &0.054\\
    1500 &1641.1 &1408.2 &-0.14 &0.11\\
    \hline
    100 &114.8  &109.2 &-0.070 &0.097\\
    700 &786.5 &706.6  &-0.11 &0.063\\
    1500 &1691.0  &1409.8  &-0.17 &0.13\\
    \hline
    \specialrule{.1em}{.05em}{.05em}
  \end{tabular}
  \caption{Comparison of simulated true electron density ($\bar{N}_{\rm e,true}$ in unit of pc\,cm$^{-3}$) and estimated one ($\bar{N}_{\rm e,est}$) for different number of lens $n_l$. The ratio is defined by Eq.\,(\ref{eq:dmratio}). $\tau$ is defined by Eq.\,(\ref{eq:tau-area}). Gaussian lens model is adopt in all the tests. In the bottom three rows, higher density gradient is used in the simulation.}
  \label{tab:table1}
\end{table}

\subsection{Luminosity function of sources}

\begin{figure}
\includegraphics[width=8.5cm]{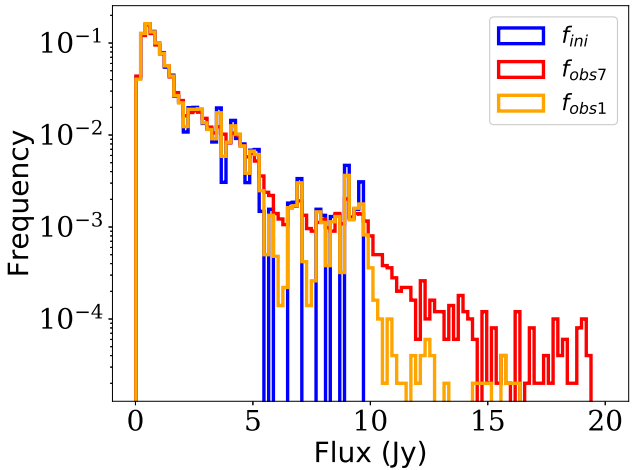}
\caption{The input luminosity function (blue) and the observed luminosity function (red, orange). To generate this figure, 700 (red) and 100 (orange) plasma lenses are placed in the field. }
\label{fig:lumi-func}
\end{figure}

As with gravitational lensing, plasma lensing can also change the luminosity function of background sources. To reiterate, plasma lensing generates both magnification and de-magnification. Since the deflection caused by plasma is relatively small, we neglect the spatial concentration by plasma lensing. The same Gaussian lens populations as in the previous section are used in this example. We randomly sample 5000 sources uniformly over the field and follow the luminosity function which is adopted from FRBcat\footnote{https://www.frbcat.org/} and the CHIME/FRB catalogue \citep{2016PASA...33...45P,2017arXiv171008155P,2021arXiv210604352T}. The observed luminosity of the source is simply calculated by $L_{\rm obs}=L_{\rm ini}\mu$. We compare the initial and observed luminosity functions in Fig.\,\ref{fig:lumi-func}. Those sources de-magnified by the lenses with luminosity below the minimum initial luminosity will be discarded as unobservable sources ($\sim 0.1-0.5\%$ for 700 lenses, $\bar{N}_{\rm e}\sim 800$ pc\,cm$^{-3}$.). The observed luminosity will have a wider range of minimum and maximum luminosity. One can see that most of the high flux sources are lensed.

%
\section{Summary and discussions}
\label{sec:summary}

Inhomogeneous distributions of plasma can deflect the propagation of low-frequency radio signals, similar to the gravitational lensing phenomena. This plasma lensing shares several similarities with gravitational lensing, especially in terms of the mathematical description. However, for an over-dense clump of plasma, it behaves analogously to a concave lens. The plasma lens thus diverges the light of background sources and produces significant de-magnification.
We study the de-magnifying properties, and the cross section for two density profiles: the Gaussian model and the softened power-law model. 

The properties of the magnification and de-magnification of Gaussian plasma lenses can be determined by the ratio of lens parameters $\theta_0/\sigma$. The cross sections of magnifications increase with this ratio. Softened power-law lenses can be described by $\theta_0/\theta_c$, and its cross sections increase with ratio as well.
Since plasma lensing causes a strong de-magnification, i.e. $\mu\rightarrow0$, there will be two observational consequences: 1) the luminosity function of the background sources is modified by plasma lensing. 2) in the study of using distant radio sources to trace the distribution of electron density in the universe, there is a bias: when the electron density is high and has a large gradient, we cannot receive any radio signal. Therefore, the average electron density that we estimate will always be lower than their true mean value. We use a toy model in a small field of view ($1\times 1$ arcmin$^2$), a reasonable electron density range from observations \citep{2017ApJ...835...29Y}, and compare the simulated and the estimated electron density. Underestimates are obtained in all our tests. The magnitude of the bias is correlated with several properties: the number of lenses, the density model, and the density parameter, e.g. $\theta_0/\sigma$ etc. In general, the higher the electron density, the larger the bias. From the observed electron density within the Milky Way, e.g. up to $\sim2000$ pc\,cm$^3$ along the disk\citep{2017ApJ...835...29Y}, the underestimate can be up to $15\%$ in our test using the Gaussian density model when the lensing probability is $\tau\approx 10\%$ (see Eq.\,\ref{eq:dmratio}).
We also calculate the luminosity function after plasma lensing, and find that the distribution of the flux of background sources can be broadened to both higher and lower values. One interesting point is that most of the high flux samples are magnified by lensing.

Our study is simplified in several aspects. First of all, the plasma lensing is frequency dependent. We adopt $\nu=1$ GHz in our study. In reality, for observations with lower frequency, plasma lensing effects will become even stronger. By comparing observations, e.g. luminosity function at different frequencies, one can obtain tighter constraints on both the plasma density and the background sources.
Another two of simplifications are the model of electron density and the luminosity function of the radio sources. The constraint on the electron density is weak from observations so far, especially on small scales. One will be able to rely on high resolution numerical simulations, which have been undergoing rapid improvement in recent years. Interests in the luminosity function have experienced explosive growths thanks to the observations of FRBs. We expect that thousands of FRBs will be detected in the coming years. The collective study of many sources affected by foreground lenses (e.g. by variations in the local electron density or by cosmic re-ionisation) will be dramatically improved to high precision. Under such conditions the magnification effect due to plasma lensing will present a systematic bias and will need to be accounted for.

\section*{Acknowledgements}
We thank the referee for a constructive and valuable report which increases the scope of our work significantly. We also thank Jenny Wagner for interesting discussions. We acknowledge use of the CHIME/FRB Public Database, provided at https://www.chime-frb.ca/ by the CHIME/FRB Collaboration. XE is supported by the NSFC Grant No. 11873006. SM is supported by the National Key Research and Development Program of China No. 2018YFA0404501, and NSFC Grant No. 11821303, 11761131004 and 11761141012. 
\section*{Data availability}
The data underlying this article will be shared on reasonable request to the corresponding author.

\bibliographystyle{mnras}
\bibliography{csection,plasmalens,new}

\label{lastpage}
\end{document}